\documentclass[11pt,a4paper]{article}
\usepackage{mathrsfs}
\usepackage{amsmath}
\usepackage{amssymb}
\usepackage{dsfont}
\usepackage{esint}
\usepackage{youngtab}

\usepackage{amsmath,amssymb,amsfonts,a4wide,graphicx,bm,times,psfrag,wrapfig,sidecap}
\usepackage{cite}
\usepackage[colorlinks=true,linkcolor=black, citecolor=black,urlcolor=black]{hyperref}
\numberwithin{equation}{section}
\makeatletter \let\old@startsection=\@startsection
\renewcommand{\@startsection}[6]
{\old@startsection{#1}{#2}{#3}{#4}{#5}{#6\mathversion{bold}}}
\makeatother

\def\Res #1{ \underset{#1}{ \rm Res}}

\def\<{\langle}
\def\>{\rangle}
\def\tr{{\rm   tr} }

\def\XXint#1#2#3{{\setbox0=\hbox{$#1{#2#3}{\int}$}
\vcenter{\hbox{$#2#3$}}\kern-.5\wd0}}

\usepackage{bm}% bold math

\begin{document}

\thispagestyle{empty}

\begin{flushright}

\end{flushright}

\vspace{1cm}
\setcounter{footnote}{0}

\begin{center}

{\Large\bf      Towards a Non-equilibrium Bethe ansatz for the    Kondo Model}

\vspace{7mm}
Eldad Bettelheim  \\[5mm]

{\it Racah Inst.  of Physics, \\Edmund J. Safra Campus,
Hebrew University of Jerusalem,\\ Jerusalem, Israel 91904 \\[5mm]}

\end{center}
\abstract{We give integral equations for the generating function  of the cummulants of the work done in a quench for the Kondo model in the thermodynamic limit. Our approach is based on an extension of the thermodynamic Bethe ansatz to non-equilibrium situations. This extension is made possible by  use of a large $N$ expansion of the overlap between  Bethe states. In particular, we make use of the Slavnov determinant formula for such overlaps, passing to a function-space representation of the Slavnov matrix . We leave the analysis of the resulting integral equations to future work.  }

\section{Introduction}
The Bethe ansatz method, once applicable, is very effective in predicting equilibrium properties of quantum integrable models. There has been a great deal of progress made since Bethe's original work on the Heisenberg model in extending the scope of and understanding the ultimate limitations of the method in thermodynamic equilibrium. The scope of possible Non-equilibrium extensions of the method are far less understood.

In this paper we attempt to obtain a non-equilibrium extension to the Bethe ansatz for the Kondo model. We consider for simplicitly one of the most basic quantities that may be considered in a quench process, a quantity which, at the same time, is, nevertheless, complex enough to be archetypical of a wider range of quantities which may considered out of equilibrium. In particular, we consider the generating function for work done in a quench. 

We identify the main problem of computing such a quantity as being the ability to utilize the Bethe ansatz to compute the overlaps between the initial state, in which the system is to be found before the quench, to any final state, at which the system may be found after the quench. We make use of the Slavnov determinant formula for such overlaps\cite{Slavnov}. The formula is useful in that it gives an expression for such overlaps, but has the drawback that in the thermodynamic limit it becomes too complex. Indeed, we are interested in obtaining a set of equations for nonequilibrium quantities, the complexity in solving of which, does not scale with $N$, where $N$ is the number of particles in the system. The Slavnov formula, on the other hand gives overlaps as $N\times N$ determinants. To overcome this difficulty we apply methods developed in Refrs. \cite{Gorohovsky:Bettelheim:Coherence:Announce,Gorohovsky:Bettelheim:A:Derivation,Kostov:Bettelheim:SemiClassical:XXX} to obtain integral equations, whose complexity does not scale with $N$, to describe the effect of the overlaps on the physical quantities we wish to compute. 

The rest of the paper is dedicated to carrying out this program. We included a more comprehensive treatment of the Bethe ansatz approach to the Kondo problem, in order for the paper to be self contained. A reader familiar with that solution may skip the pertinent sections, after reading such advice included at the beginning of the respective section.  

We wish to draw the reader's attention to the following notation, which shall be used throughout: the imaginary unit will be denoted by $\imath=\sqrt{-1}$, not to be confused with the letter $i$, which will often be used as an index. 

\section{Model}
We consider the Kondo model 
\begin{align}
H=\int -\imath\psi^\dagger_{\sigma}(x)\partial_x \psi_{\sigma}(x)  +\frac{I}{2}  \psi_\sigma^\dagger(0)\vec{\sigma}_{\sigma \sigma'}\psi_{\sigma'}(0)\cdot \vec{S,}
\end{align}
with the summation of repeated indices being implied, and set:\begin{align}
g_I=\tan\left(\frac{7}{4} I\right).
\end{align}

A Bethe wave function for $N$\ electrons  is  characterized by the magnon number $M$ (number of down spins) , a set of $M$\ magnon  rapidities $\bm{\lambda},$ and a set of $N+1$\ wave numbers $\bm{k}$. We describe the wave function in a non-standard fashion. Let $\tilde\psi_{\bm\sigma \otimes s}(\bm x)$ be the {\it\ standard} representation of the wave function, where $s$\ is the spin component of the impurity, $\bm \sigma $ is the vector of spin components of the electrons and $\bm x$ is the vector of electron positions. The vector ${\bm\sigma \otimes s} $ is  given by $(\sigma_1, \sigma_2, \dots, \sigma_N, s).    $The function  $\tilde\psi_{\bm\sigma \otimes s}(\bm x)$ gives the probability amplitude to find first  electron at position $x_1$ with spin component $\sigma_1$, the second electron at position $x_2$ with spin component $\sigma_2$, etc., and the impurity with spin component $s$.    We shall be using, following Refrs.  \cite{AndreiL:First:Kondo,Wiegmann:First:Kondo}, a non-standard representation,  namely we will describe the same quantum state by the function $\psi_{\bm\sigma \times s}(\bm x)$, which is related to $\tilde \psi$ by:
\begin{align}
\label{tildedef}\psi_{Q\bm\sigma\otimes s}(\bm x)=\tilde{\psi}_{\bm\sigma\otimes s}(\bm x ),
\end{align}
where $Q$ is a permutation which orders the $x_i$'s:  $x_{Q(1)}<x_{Q(2)}<\dots<x_{Q(N)}$. In other words, the vector $Q\bm x$ is ordered (we define the action of a permutation on a vector in a natural way: $(Q\bm v)_i = v_{Q(i)}$) . The function $\psi_{\bm\sigma\otimes s}(\bm x)$ then gives the probability amplitude to find electrons at $x_1,x_2,\dots,x_N$ in such a configuration as that their spin components, if we inspect the electrons from left to right, are given by $\sigma_1,\sigma_2,\dots,\sigma_N$, while the impurity has spin component $s$.  

The eigenstates in this non-standard representation have a nested Bethe ansatz form:
\begin{align}
\psi_{\bm \sigma\otimes s}(\bm x) = \sum_{P\in S_N} {\rm sign}(P)\Psi_{P\circ Q}(\bm \sigma\otimes s)e^{\imath \bm x \cdot P \bm k},
\end{align}
where $Q$ is, as before, the permutation that orders $\bm x$.   Crucially,  due to the peculiarity of the Kondo Bethe ansatz (a different situation can be found, e.g., in the Hubbard model), $\Psi_P$ turns out not to depend on $P$ and the wave function separates:
\begin{align}
\psi_{\bm \sigma}(\bm x) =\label{seperatedWF}\left(\det_{i,j}e^{\imath k_{i} x_{j}}\right)\Psi(\bm \sigma\otimes s) ,
\end{align}
where here it is no longer assumed that the $x_i$'s are ordered. The first factor is a function of $x$ only,  while $\bm{k}$ is a vector of $N$\ quantum numbers, or wave number rapidities. The second factor depends only on the  spin configuration, where $\bm{\lambda}$ is  a vector of $M$\ quantum numbers, or magnon rapidities. The number $M$ is equal to the number of down spins. 

Note, however, that the separation occurs only in the non-standard basis after applying an ordering transformation that acts on the spin components. Such a separation in the  non-stadard representation can  only occur in cases where  the kinetic energy is linear in  momentum. Indeed, the wave  function in the standard representation, $\tilde{\psi}$, which corresponds to (\ref{seperatedWF}), does not have a continuous derivative with respect to any of the variables $x_i$ across the hyper-plane $x_j=x_k$. Nevertheless, since the Hamiltonian is first order in derivative,  continuity of the derivative is not required. 

It is important to note that even though we are writing the wave function in the representation where the vector of spin components, $\bm\sigma,$  is ordered (namely the spin components are measured from the left most particle to the rightmost particle), the inner product remains the standard one:
\begin{align}
\<\psi^{(2)}|\psi^{(1)}\> =\sum_{\bm\sigma} \int \psi_{\bm \sigma}^{*(2)}(\bm x)\psi_{\bm \sigma}^{(1)}(\bm x)d^N{\bm x}  =\sum_{\bm\sigma} \int \tilde\psi_{\bm \sigma}^{*(2)}(\bm x)\tilde\psi_{\bm \sigma}^{(1)}(\bm x)d^N{\bm x}, 
\end{align} 
which is easily shown by passing to a sum over $Q\bm\sigma $ in the expression after the first equality and using (\ref{ThetaDEf}). 

The factorization property may be ascertained by examining the explicit form of $\Psi,$ which is known from Yang and Yang's work \cite{Yang:Yang:1966:I:Heisenberg,Yang:Yang:1966:II:Heisenberg,Yang:Yang:1966:III:Heisenberg,Yang:1967:PRL:Delta:Function,Yang:Yang:Bosons:With:Delta:1969} and its application to the Kondo problem\cite{AndreiL:First:Kondo}, namely we have
\begin{align}
&\Psi(\bm \sigma \otimes s) =\sum_{ P\in S_M} A(P\bm{\lambda}) \prod_{l=1}^M F[(P\bm{\lambda})_l,y_l]\label{BigPsi}
\end{align}
where :
\begin{align}
&\label{BigF}F[{\lambda},y] =\left(\frac{\lambda- \imath/2}{\lambda  +\frac{1}{g} -  \imath /2}  \right)^{\delta_{y,N+1}}\left(\frac{\lambda+ \imath/2}{\lambda- \imath/2}\right)^{y-1},\\
&A(\bm{\lambda})=\prod_{1\leq m <n \leq M} \frac{\lambda_m-\lambda_n-\imath}{\lambda_m-\lambda_n}
\end{align}
where $y_l$ is the $l$'th down spin in the sequence $\bm{\sigma}\otimes s$.  

We shall also use Dirac bra-cket notations, in which the factorization property reads:
\begin{align}
&|\bm{\lambda},\bm{k}\> =|\bm{\lambda}\>\otimes|\bm{k}\>\\& 
\<\bm{x}|\bm{k}\>=\det_{i,j}\left[  e^{\imath k_i\cdot  x_j}\right]\label{ChargeWF}\\
&\<\bm{\sigma}\otimes s|\bm{\lambda}\>=\varphi(\bm \sigma \otimes s)
\end{align}

\subsection{ Bethe Ansatz Equations}

If the rapidities $\bm \lambda$ and $\bm k$ satisfy:
\begin{align}
&\label{BetaLambda}
\left(\frac{\lambda_l  + \imath/2}{\lambda_l  - \imath/2}\right)^N \frac{\lambda_l  +\frac{1}{g}+ \imath /2}{\lambda_l  +\frac{1}{g} -  \imath /2}=- \prod_{m=1 }^M\frac{\lambda_l  - \lambda_m+ \imath}{\lambda_l  - \lambda_m- \imath},\\
&\label{betaK}e^{\imath k_j L }=e^{\imath I/4}\prod_{l=1}^M\frac{\lambda_l  + \imath /2}{\lambda_l  - \imath /2}\equiv e^{\imath \delta }
% \\&e^{ik_{N+1} L }        =V^{-N}\prod_{l=1}^M\left(\frac{\lambda_l  - %\frac{1}{2}- i U/4}{\lambda_l  - \frac{1}{2}+ i U/4}\right)^N\frac{\lambda_l % - i U/4}{\lambda_l  + i U/4}
\end{align} 
then $|\bm{\lambda},\bm{k}\>$ is an eigenstate of $H$.

To describe a solution of (\ref{BetaLambda}) first take the logarithm of that equation:
\begin{align}
\label{BetheLog}N\log\left(\frac{\lambda_l+\frac{\imath}{2}}{\lambda_l-\frac{\imath}{2}}\right) +\log\left(\frac{\lambda_l  +\frac{1}{g}+\frac{\imath}{2}}{\lambda_l  +\frac{1}{g} -\frac{\imath}{2}}\right)= 2\pi \imath n + \sum_{m=1}^M \log\left(\frac{\lambda_l-\lambda_m + \imath}{\lambda_l-\lambda_m - \imath}\right).
\end{align}
The number $n$ on the RHS which arises because of the ambiguity of taking the logarithm, is called 'the mode number' and we thus may assign a mode number for any root $\lambda_l$ label by a given $l$. The roots arrange themselves in $m$- strings, namely, a set of $m$ Bethe roots $\lambda_{l_0}$, $\lambda_{l_0+1},\dots,$ $\lambda_{l_0+m-1}$ of the form:
\begin{align}
\label{m-string}\lambda_{l_0+j}=\alpha_{i_{l_0}}^{(m)}+\frac{\imath}{2} (2j-m+1), 
\end{align}
 where $\alpha_i^{(m)}$\ is a real number. 
A $1$-string is a simple root on the real axis. The $\alpha_{i}^{(m)} $ are ordered in ascending order, $\alpha_{l}^{(m)}< \alpha_{l+1}^{(m)}$ .
All rapidities in a given string share the same mode number, $n$$.$ Rapidities which belong to different  strings of different length can share the same mode number but not if the rapidities belong to different strings of equal length.
 
Since all rapidities in a given string share the same mode number, we can assign for each $\alpha_l^{(m)}$ a mode number $n(l,m)$, defining  a function $n(l,m)$. We can define the density of strings of length $m$ as  
\begin{align}
\sigma^{(m)}(\alpha^{(m)}_l) = \frac{1}{\alpha^{(m)}_{l+1}-\alpha^{(m)}_l},\label{sigmaMDef}
\end{align}
 and the density of 'holes' in the distribution of $m$-strings: 
\begin{align}
\sigma^{(m)}_h(\alpha^{(m)}_l) =\left[n(l+1,m)-n(\l,m)-1\right]\sigma^{(m)}(\alpha^{(m)}_l).\label{sigmaMhDef}
\end{align}
In the continuum limit the  strings become dense and we can interpolate  $\sigma^{(m)}(\alpha^{(m)}_l)$ into a smooth  function $\sigma^{(m)}(\lambda),$  the meaning of which is the density of strings of length $m$ around point $\lambda$. The function  $\sigma^{(m)}_h(\lambda)$, on the other hand, gives the density of $m$-holes around $\lambda$, namely the density of points at which, due to the availability of an unused mode number, an $m$-string could potentially have been placed, but has not.    

We may treat $\sigma_h^{(m)}$ as a free parameter, while $\sigma^{(m)}$ is determined by it, through the Bethe ansatz equations (\ref{BetaLambda}). Let us write ${\bm \sigma}_h$ for the whole set of functions $\{\sigma^{(m)}_h\}_{m=1}^\infty$ and ${\bm \sigma}$ for the  set of functions $\{\sigma^{(m)}_h\}_{m=1}^\infty.$ 

% Finally, we write down another useful definition that will appear repeatedly in the following:
% \begin{align}
% \label{EpsilonDef}\varepsilon_m(\lambda) =T\log\left[\frac{\sigma^{(m)}(\lambda)}{\sigma_h^{(m)}(\lambda)}\right],
% \end{align}
% where $T$ at the moment serves as an arbitrary parameter, to be given a meaning later. The importance of $\varepsilon_m(\lambda)$ lies in the fact that, as we shall see,  at equilibrium, although $\sigma^{(m)}(\lambda)$ and $\sigma_h^{(m)}(\lambda)$ depend in important ways on the existence of an impurity, $\varepsilon_m(\lambda)$ does not, and takes the same form as for free fermions. 

\section{Loschmidt Echo}

The method described in this section was introduced in \cite{Caux:Essler:Time:Evolution:After:Quench} and applied in Refrs. \cite{Caux:Lieb:Liniger:Quench,Caux:Heisenberg:Ising:Quench,Essler:Bertini}, it was dubbed 'the quench action approach'. 
We may wish to compute the Loschmidt echo, $L,$ given by the following:
\begin{align}
&\label{Echo}L(t) = \sum_{k} e^{-\gamma E_k} \left|\<i | E_k\>\right|^2,
\end{align}
where $|i\>$ is some initial state of the free fermion system (without the impurity). We shall assume for simplicity that this is an eigenstate of the free Fermion problem.  The Loschmidt echo is the generating function of the commulants $C_n = \<\Delta w^n\>_c$ of the work per electron done on  the system, $\Delta w$, after time $t$ due to a quench:
\begin{align} 
\mathcal{W}(\gamma)=\frac{1}{N} \log L(\gamma) = \sum_n  \frac{C_n (-\gamma)^ n}{n!}.
\end{align} 

It is customary to compute the Loschmidt echo for imaginary $\gamma$, but as a generating function for the cummulants, and in the method we take to compute it, it will prove below more convenient to take $\gamma$ to be real. 

Suppose, as we shall see later, we know how to compute $\left|\<i| E_k\>\right|$ given $\bm{\sigma}$. We wish to compute $L(t)$ armed with this knowledge . To simplify notations we shall write:

\begin{align}
\mathcal{\label{ADef}A}({\bm \sigma})= \log\left|\<i | E_k\>\right|^2.
\end{align}

 One may represent the Loschmidt echo as follows\cite{Caux:Essler:Time:Evolution:After:Quench}:
\begin{align}
L(t)=\int \exp\left[S-\gamma E+  \mathcal{A} \right]  \mathcal{D}{\bm \sigma}.
\end{align}
where $e^S$ is the measure of integration when passing from the sum over all possible configurations of solutions of the Bethe equations, to the variables ${\bm \sigma}$.

The measure of integration $e^S$ appears already in the thermodynamic Bethe ansatz, and has a well defined expression appropriate for an entropy, to be discussed below. The quantities $E, S, $ and $\mathcal{A  }$  are all functionals of ${\bm \sigma}_h$. Furthermore, $E$  and $S$ and $\mathcal A$ are extensive quantities, and accordingly we may compute $\frac{1}{N}\log L(t)$ by saddle point, the result becoming exact in the thermodynamic limit. 

To facilitate writing the thermodynamic limit, we define:
\begin{align}
T\equiv\frac{1}{\gamma}
\end{align}
and:\begin{align}
M(T,\bm\sigma) \equiv E(\bm\sigma)- T (S(\bm\sigma)+\mathcal{A}(\bm\sigma)
)=F(\bm\sigma) - T\mathcal{A}(\bm\sigma)
\end{align}
The saddle point equations for $M(T,\bm\sigma)$ are:
\begin{align}
0=\left.\frac{\delta M}{\delta \bm \sigma}\right|_{\bm \sigma = \bm\sigma^* },
\end{align}
which can be also expressed as:
\begin{align}
\left.\frac{\delta F}{\delta \bm \sigma}\right|_{\bm \sigma = \bm\sigma^* }=T\left.\frac{\delta \mathcal A}{\delta \bm \sigma}\right|_{\bm \sigma = \bm\sigma^* }.\label{VaraiationOfFisOfA}
\end{align}
We shall write this equation in terms of a set of integral equations for $\bm \sigma^*$. Once  $\bm \sigma^*$ is found, for example by solving these equations numerically, the cummulant generating function may be computed: 
\begin{align}
\mathcal{W}(\gamma) =\frac{\gamma}{N} M(\gamma^{-1},\bm \sigma^*).
\end{align}
It will be more useful in fact to write down the following differential equation for the generating function:
\begin{align}
\frac{\partial  }{\partial \gamma} \mathcal W(\gamma) =\frac{1}{N}   E(\bm \sigma^*),
\end{align}
which requires only computing the energy associated with $\bm \sigma^*$ and avoids computing $\mathcal{~A}(\bm \sigma^*). $ Indeed, we shall never give an  equation for $\mathcal{A}$ itself but only for its derivative with respect to $\bm \sigma$.

We shall refer to the right hand side of equation (\ref{VaraiationOfFisOfA}), as 'non-equilibrium source'. The main challenge in applying the Bethe ansatz to non-equilibrium situations is the task of computing non-equilibrium sources. The current paper's main objective is developing a method to compute them. We shall obtain linear integral equations
determining the non-equilibrium source. We note that Eq. (\ref{VaraiationOfFisOfA}) itself, with sources set to zero, can be brought to a form of a   set of nonlinear integral equations familiar from the equilibrium Thermodynamic Bethe ansatz. The introduction of a non-equilibrium source, then supplements these integral equations by further linear integral equations. These latter equations depend, themselves, on the densities $\bm \sigma$, $\bm \sigma_h$ nonlinearly. 

The method described in this section was introduced in \cite{Caux:Essler:Time:Evolution:After:Quench} and applied in Refrs. \cite{Caux:Lieb:Liniger:Quench,Caux:Heisenberg:Ising:Quench,Essler:Bertini}, where techniques somewhat different than what will be described in the sequel, and thus of a different scope of application, have been used to compute the non-equilibrium sources. 

\section{Form of Bethe Ansatz Eigenstates}
In order to obtain the non-equilibrium sources in question, we must have a better grasp of the form of the eigenstates for the Kondo Bethe ansatz. This section follows closely \cite{Andrei:Lowenstein:Review,Wiegmann:Tsvelick:Review:Kondo} and may be skipped by a reader familiar with the Bethe ansatz for the Kondo model. The method is based on the application of the Bethe ansatz to the Heisenberg chain as described originally in \cite{Gaudin:Heisheberg:Thermo,Takahashi:Heisenberg:Finite:T,Takahashi:Kondo:Numerics,Takahashi:Suzuki:Heisheberg:Anisotropic:Finite:T,Johnson:McCoy:Heisenberg:Low:Temp}.

\subsection{Ground State}We first study the ground state of the Kondo problem. The ground state may be found by taking the  continuum limit of (\ref{BetheLog}), which reads:
\begin{align}
\label{ContinuumBethe}N\log\left(\frac{\lambda_l+\frac{\imath}{2}}{\lambda_l-\frac{\imath}{2}}\right) +\log\left(\frac{\lambda_l  +\frac{1}{g}+\frac{\imath}{2}}{\lambda_l  +\frac{1}{g} -\frac{\imath}{2}}\right)= 2\pi \imath n + \int d\lambda' \sigma(\lambda') \log\left(\frac{\lambda_l-\lambda' + \imath}{\lambda_l-\lambda' - \imath}\right)
\end{align}
where the density $\sigma(\lambda_l) = \frac{1}{\lambda_{l+1}-\lambda_l}$, has already been defined in (\ref{sigmaMDef}), by setting $m=1$.

 One may now subtract equation  (\ref{ContinuumBethe}) for $\lambda_{l+1}$ from the same equation for $\lambda_l$ to obtains:
\begin{align}
2\pi\left[\sigma^{(1)} (\lambda)+\sigma^{(1)}_h(\lambda) \right]\label{BasicBethe}=\frac{     N}{\lambda^2+\frac{1}{4}} +\frac{ 1}{(\lambda+\frac{1}{g})^2+\frac{1}{4}}-  \int d\lambda' \frac{2\sigma^{(1)}(\lambda')}{(\lambda-\lambda')^2+1}, 
\end{align}
where it is assumed that there are no strings of length larger than $1$ in the ground state. In fact, the ground also contains no holes, so we may also set $\sigma^{(1)}_h(\lambda)=0.  $  These two assumptions (that the ground state has no holes and no string of length larger than $1$)  need to be ascertained by proving that all other states have higher energy. We do not perform this check here, but only note that   Eq. (\ref{BasicBethe}) may be solved to give the ground state density of rapidities: 

%  To find the ground state density $\sigma(\lambda)$, use the fact that:
% \begin{align}
% &\int_{-\infty}^\infty\frac{d\lambda'}{1+(\lambda-\lambda')^2}\frac{2}{\cosh(\pi\lambda') } = \frac{2\pi}{\cosh[\pi(\lambda+\imath)]} -4\sum_{n=0}^\infty \frac{(-)^n}{1+(\lambda-(n+\frac{1}{2})\imath)^2}=\nonumber\label{coshofm=2} \\
% &=-\frac{2\pi}{\cosh(\pi\lambda)} -2\imath \sum_{n=0}^\infty \left(\frac{(-)^n}{\lambda-(n+\frac{3}{2})\imath} -\frac{(-)^{n-2}}{\lambda-(n-\frac{1}{2})\imath}  \right)=\nonumber \\&=-\frac{2\pi}{\cosh(\pi\lambda)}+\frac{2}{\lambda^2 +\frac{1}{4}},
% \end{align}
% to obtain:
\begin{align}
\sigma^{(1)}_{gs}(\lambda) = \frac{N}{2\cosh\pi \lambda} + \frac{1}{2\cosh \pi(\lambda+\frac{1}{g})},\label{KondoGS}
\end{align}
all other $\sigma^{(m)}, \sigma_h^{(m)}$ being equal to zero in the ground state.

\subsection{Excitations}
As mentioned above, an $m$-string is a set of $m$ Bethe roots $\lambda_{l_0}$, $\lambda_{l_0+1},\dots,$ $\lambda_{l_0+m-1}$ of the form:
\begin{align}
\label{m-string}\lambda_{l_0+j}=\alpha^{(m)}+\frac{\imath}{2} (2j-m+1). 
\end{align}
 Note again that strings of different length can share the same mode number but not strings of equal length. 

To obtain and equation for $\sigma^{(m)}$ consider Eq. (\ref{BetheLog}), which are the Bethe equations in logarithmic form. These equations can be cast as  equation for the $\alpha_i^{(m)}$'s only, as the position of all rapidities are known  if the position of the midpoint of each string, $\alpha_i^{(m)}$ are known. Such an equation can be obtained by going over all strings and performing the sum  (\ref{BetheLog})  over all rapidities belonging to each  string. One obtains:
\begin{align}
&\label{StringBetheNoDeriv}N \Theta_m(\alpha_i^{(m)}) + \Theta_m\left(\alpha_i^{(m)}+\frac{1}{g}\right) =\\&=\nonumber 2\pi \left( n(i,m)+\frac{m+1}{2}\right)-\imath \sum_{m',j}q_{m,m'}\left(\alpha_i^{(m)}-\alpha^{(m')}_j\right),
\end{align}
where 
\begin{align}
\label{ThetaDEf}\Theta_a(\lambda)=-\imath \log\frac{\lambda+\frac{\imath}{2} a}{\lambda-\frac{\imath}{2} a},
\end{align}
and\begin{align}
&\label{qExpression}q_{m,m'}(\lambda)=\sum_{j=0}^{m-1} \imath \Theta_{m+m'-2j}\left(\lambda\right)+\imath \Theta_{m+m'-2(j+1)}\left(\lambda\right).
\end{align}

We may subtract the equation for $i+1$ from the equation for $i$  (keeping $m$ fixed) to obtain:
\begin{align}
\label{FinalBetheStrings} \sigma_h ^{(m)}(\lambda)+  \int d\lambda' K_{m,m'}(\lambda-\lambda')\sigma^{(m')}(\lambda')=NG_m(\lambda) +G_m(\lambda+g^{-1}),
\end{align}
where 
\begin{align}
&\label{GDef}G_m(\lambda)=\frac{1}{2\pi}\frac{     Nm}{\lambda^2+\frac{m^2}{4}}, \\
&\label{KDef}K_{m,m'}(\lambda)=\delta_{m,m'}\delta(\lambda)+ \sum_{j=0}^{\min(m,m')-1}G_{m+m'-2j}(\lambda)+G_{m+m'-2(j+1)}(\lambda),
\end{align}
and the definition of $\sigma_h^{(m)}$ is given in (\ref{sigmaMhDef}). 

To solve (\ref{FinalBetheStrings}) for $\sigma^{(m)}$ given $\sigma_h^{(m)}$ for all $m$,  we may apply the Fourier transform. We first Fourier transform $G_m$, the transform being denoted by $\tilde G_m$:
\begin{align}
&\tilde{G}_m(\omega)= \int e^{-\imath \omega \lambda} G_{m}(\lambda) d\lambda=e^{-\frac{m}{2}|\omega|}-\delta_{m,0}\label{GmTransform}
\end{align}
The Fourier transform of $K_{m,m'}$, which we denote by $\tilde K_{m,m'}$ is given by
\begin{align}
&\tilde{K}_{m,m'}(\omega)=2e^{-\frac{1}{2}|\omega|\max(m,m')} \coth\left({|\omega|/2}\right)\sinh\left( {\frac{1}{2}|   \omega|\min(m,m')}\right) 
\end{align}
Note that from this equation, by manipulations involving only simple hyperbolic-trigonometric identities, one may derive:
\begin{align}
\label{Resolvent1}&\tilde K_{m+1,m'}+\tilde K_{m-1,m'} =2\cosh\left({|w|/2}\right)\tilde K_{m,m'} -2\cosh\left({|w|/2}\right)\delta_{m,m'}
\end{align}
This shows that, treating $\tilde K_{m,m'}(\omega)$ as a matrix with indices $m$ and $m'$ can be inverted as follows:
\begin{align}
\label{Resolvent2}\left(\mathds{}\tilde K(\omega) \right)^{-1}_{m,n}=-\frac{1}{2}\cosh^{-1}\left({|w|/2}\right)\left( \delta_{m,n+1}+\delta_{m,n-1}\right)+\delta_{m,n}.
\end{align} 
Indeed applying the RHS of  (\ref{Resolvent2}) to $\mathds{}\tilde K,$ and making use of (\ref{Resolvent1}), the identity becomes apparent. Fourier transforming back to $\lambda$ space yields the following form for the inverse (as a kernel) of $K$:
\begin{align}
\label{KInverse}(K)^{-1}_{m,n}(\lambda) = \delta_{m,n}\delta(\lambda) -\frac{\delta_{m,n+1}+\delta_{m,n-1}}{2\cosh
\pi\lambda} .
\end{align}

% It is useful at this point to examine the algebra of the kernels $G$ and $K$ with respect to convolution, $(a\star b)(\lambda) \equiv   \int a(\lambda-\lambda')b(\lambda')d\lambda'$. Indeed, 
% \begin{align}
% &G_m\star G_n =\frac{1}{(2\pi)^2} \int \frac{m}{(\lambda-\lambda')^2+\frac{m^2}{4} }\frac{n}{\lambda'^2+\frac{n^2}{4} } d\lambda'= \\
% &=\frac{1}{2\pi}\left[ \frac{m}{(\lambda-\imath \frac{n}{2})^2+\frac{m^2}{4}} +\frac{n}{(\lambda+\imath \frac{m}{2})^2-\frac{n^2}{4}} \right] = G_{m+n} .
% \end{align}
% Thus, in particular, the $G$'s commute under convolution. 

% We may also define the kernel $s$ by
% \begin{align}
% s(\lambda) = \frac{1}{\cosh(\pi \lambda).}
% \end{align}
% And the action of $s$ on $G_m$ may be computed as follows:
% \begin{align}
% &s\star G_m =\frac{1}{2\pi} \int \frac{d\lambda'}{\cosh\left[\pi(\lambda-\lambda')\right]}  \frac{m}{\lambda'^2+\frac{m^2}{4}}  = \\
%  &= \cosh^{-1}\left[\pi \left( \lambda-\frac{\imath m}{2}\right) \right] +\imat \sum_{n=0}^\infty (-)^n \frac{m}{(\lambda-\imath\frac{2 n +1}{2})^2+\frac{m^2}{4}} = \\ 
% &= \cosh^{-1}\left[\pi \left( \lambda-\frac{\imath m}{2}\right) \right] - \sum_{n=0}^\infty (-)^n \left[\frac{1}{\lambda-\imath\frac{2 n +1 +m}{2}} -\frac{1}{\lambda-\imath\frac{2 n +1 -m}{2}}\right] = \\
% &=\cosh^{-1}\left[\pi \left( \lambda-\frac{\imath m}{2}\right) \right] +\cosh^{-1}\left[ \pi\left( \lambda + \frac{\imath m}{2}\right)\right] 
% \end{align}

Acting with $K^{-1}$ on (\ref{FinalBetheStrings}) one obtains the desired equation giving $\bm \sigma$ for known $\bm \sigma_h$:
\begin{align}
\label{FinalFinalBethe}\sigma^{(m)}(\lambda) &=  -\sigma_h^{(m)}\nonumber (\lambda)+ \int\frac{\sigma^{(m+1)}_h(\lambda')+\sigma^{(m-1)}_h(\lambda')}{2\cosh\left[\pi(\lambda-\lambda') \right]}\\&+\delta_{m,1}\left(\frac{N}{2\cosh[\pi\lambda]}+\frac{1}{2\cosh[\pi(\lambda+\frac{1}{g})]}\right).
\end{align}
Here we use the following  when applying   $K^{-1}$ to $G_m$:
\begin{align}
\label{InverseKonG}\sum_{m'}\int K^{-1}_{m,m'}(\lambda - \lambda')G_{m'}(\lambda') d\lambda' = \frac{\delta_{m,1}}{2\cosh(\pi \lambda)},
\end{align} 
which is most conveniently proven by using the Fourier transforms
(\ref{GmTransform}) and (\ref{Resolvent2}).

\section{Thermodynamics}
In this section we discuss the thermodynamics of the Kondo model, again following closely \cite{Andrei:Lowenstein:Review,Wiegmann:Tsvelick:Review:Kondo}, making this section to anyone who is familiar with the work reviewed in these references superfluous. The aim of this section is to find the free energy and in particular the variation of which with respect to the density of $m$-strings and to establish the different scales characteristic of these density distributions. 

 We shall denote from now on the set of string densities $\sigma^{(m)}(\lambda)$ by $\bm \sigma$. 

\subsection{Variation of the Free Energy }To find the energy of the Bethe eigenstate which is obtained by solving (\ref{FinalBetheStrings}), one must compute $k_i$ through (\ref{betaK}), and write the result through the variables $\alpha_l^{(m)}$, by taking the product over the rapidities belonging to each string. One obtains:
\begin{align}
e^{\imath k_jL} = e^{\imath I/4} \prod_{m,l} \prod_{j=0}^{m-1} \frac{\alpha_l^{(m)}+\frac{\imath}{2} (2j-m+2) }{\alpha_l^{(m)}+\frac{\imath}{2} (2j-m) }= e^{\imath I/4} \prod_{m,l}  \frac{\alpha_l^{(m)}+\frac{\imath m}{2} }{\alpha_l^{(m)}-\frac{\imath m}{2} }
\end{align}
Such that after taking the logarithm one obtains:
\begin{align}
Lk_j \label{kIsnplusdelta}= 2\pi n_j +\delta(\bm \sigma),
\end{align}
where
\begin{align}
\delta(\bm \sigma)=  I/4+ \sum_{m>0} \int \Theta_m(\lambda)\sigma^{(m)}(\lambda)d\lambda.\label{phaseShift}
\end{align}
Summing $E=\sum_j\hbar  v_F k_j$ such that summing over all particles one obtains:
\begin{align}
E &=\sum_j \frac{2\pi \hbar v  _Fn_j}{L} +E_s(\bm\sigma),\end{align}
where $\Theta_m$ is given in (\ref{ThetaDEf}) 
and
\begin{align}
E_s(\bm\sigma)=\frac{\epsilon_F I}{\pi 4}+ \frac{\epsilon_F}{\pi}  \sum_{m>0}\int \Theta_m(\lambda)\sigma^{(m)}(\lambda)d\lambda,
\end{align}
with
\begin{align} 
\label{eFDef}\epsilon_F = \frac{\pi \hbar Nv_F }{L},     
\end{align}

The entropy is given by considering all counting all possible configurations which have the same coarse grained densities $\bm \sigma$ and $\bm \sigma_h$. The cominatorics of the counting procedure yields an expression which contain factorials, which upon use of the Sterling formula allows one to write:
\begin{align}
S=\sum_m  \int\left(\sigma^{(m)}+\sigma_h^{(m)}\right)\log\left( \sigma^{(m)}+\sigma_h^{(m)}\right)- \sigma^{(m)}\log\left( \sigma^{(m)}\right)-\sigma_h^{(m)}\log\left( \sigma_h^{(m)}\right).
\end{align}
The free energy is, of course, $F=E-TS$. 

We need the variation of the free energy with respect to  the variables $\sigma_h^{(m)}, $  or $\sigma_h$ which are  free parameters. In varying $\sigma^{(m)}$ with respect to $\sigma_h^{(m)}$, one notes from (\ref{FinalFinalBethe}) 
\begin{align}
\frac{\delta \sigma^{(m)}(\lambda)}{\delta \sigma^{(m')}_h(\lambda')}  =-K^{-1}_{m,m'}(\lambda-\lambda') .\end{align}
Alternatively one may write:
\begin{align}
\frac{\delta \sigma^{(m)}_h(\lambda)}{\delta \sigma^{(m')}(\lambda')} = -K_{m,m'}(\lambda-\lambda'),
\end{align}
which may also be  directly obtained from (\ref{FinalBetheStrings}). It is more convenient, then, to take a variation with respect to $\sigma^{(m)}$ rather than with respect to $\sigma_h^{(m)}$, since the energy (if not the free energy), is written through $\sigma^{(m)}$ directly. One obtains:
\begin{align}
&\frac{\delta F}{\delta \sigma^{(m)}(\lambda)} =\frac{\epsilon_F}{\pi } \Theta_m(\lambda)-T\log\left(1+e^{-\beta{\varepsilon_{m}(\lambda)} }\right)+
\nonumber\\&+T\sum_{m'}\int K_{m,m'}(\lambda-\lambda')\log\left(1+e^{\beta\varepsilon_{m'}(\lambda') }\right),\label{dFdSigma}
\end{align}
where $\varepsilon_m(\lambda) $ is given in (\ref{EpsilonDef}).
\begin{align}
\label{EpsilonDef}\varepsilon_m(\lambda) =T\log\left[\frac{\sigma^{(m)}(\lambda)}{\sigma_h^{(m)}(\lambda)}\right].
\end{align}
We may write the derivative of the free energy with respect to  $\sigma_h^{(m)}(\lambda)$  by applying  $K^{-1}$, ($K^{-1}$ being given in (\ref{KInverse})) to obtain:
\begin{align}
\label{MainEquiEquation}& \frac{\delta F}{\delta \sigma_h^{(m)}(\lambda)}=\frac{ 2\epsilon_F}{\pi }\delta_{m,1}\tan^{-1} e^{\pi\lambda }+\varepsilon_m(\lambda)-\\
&-T      \int\frac{ \log\left[\left(1+e^{-\beta \varepsilon_{m+1}(\lambda') }\right) \left(1+e^{-\beta \varepsilon_{m-1}(\lambda') }\right)\right]}{2\cosh\left[\pi(\lambda-\lambda')\right]}
\nonumber
\end{align}
The first term on the right hand side is the result of the action of  $K^{-1}$\ on $\Theta_m$. To see that this produces the expression in question, consider that $\Theta_m'(\lambda)=- 2\pi G_m(\lambda)$ and the identity given in equation (\ref{InverseKonG}). The latter  two facts  lead immediately to   
\begin{align}
\sum_{m'}\int K^{-1}_{m,m'}(\lambda-\lambda')\Theta'_{m'}(\lambda')= -\pi \delta_{m,1}\cosh^{-1}\left(\pi \lambda \right).
\end{align} 
Integrating  by parts and shifting the derivative from $\lambda'$ to $\lambda$, one then obtains:
\begin{align}
\int K^{-1}_{m,m'}(\lambda-\lambda')\Theta_m(\lambda')=-\pi \delta_{m,1}\int^\lambda\cosh^{-1}\left(\pi \tilde \lambda \right)d\tilde \lambda=-2  \delta_{m,1}\tan^{-1} e^{\pi\lambda }.
\end{align} 
The equation must be supplemented with the asymptotics of $\varepsilon_m$ as $m\to\infty$. This is given by:
\begin{align}
\lim_{m\to\infty}\frac{\varepsilon_m(\lambda)}{m} = 0.
\end{align}
In fact, to obtain this identity it is most useful to add a magnetic field and take it to zero at the end of the calculation. We do not show this procedure here, referring the interested reader to \cite{AndreiL:First:Kondo,Wiegmann:First:Kondo} or the respective reviews \cite{Andrei:Lowenstein:Review,Wiegmann:Tsvelick:Review:Kondo}.

In equilibrium we set the left hand side of (\ref{MainEquiEquation}) to zero and solve this integral equation for $\varepsilon_m$. A knowledge of $\varepsilon_m$ is sufficient in equilibrium to derive the free energy, and hence all  properties which are considered  in equilibrium. In the non-equilibrium situation at hand, we set $T=\frac{1}{\gamma}$  and the left hand side of Eq. (\ref{MainEquiEquation})  to the non-equilibrium source as in Eq. (\ref{VaraiationOfFisOfA}). This non-equilibrium source  may then be computed along the lines given in the remainder of this paper. Such a computation requires knowledge of $\bm \sigma$ and $\bm \sigma_h$ separately, rather than just the combination $\varepsilon_m$. Namely,  $\bm \sigma$ and $\bm \sigma_h$  have to be found given a solution of (\ref{MainEquiEquation}) for $\varepsilon_m$. This task  can be achieved by   solving the linear integral equation,  (\ref{FinalFinalBethe}) for $\bm \sigma_h$, by substituting $ \sigma^{(m)}=  \sigma^{(m)}_h e^{\gamma \varepsilon_m},  $ or, alternatively,  by solving (\ref{FinalBetheStrings}) for $\bm \sigma,$ by substituting $ \sigma_h^{(m)}=  \sigma^{(m)} e^{-\gamma \varepsilon_m} $.  Both options represent additional linear integral equations, which are normally not considered in equilibrium, even though they are meaningful even in equilibrium.

\subsection{Scales\label{ScalesSubSection}} 
In this subsection we shall discuss the different scales that characterize the $m$-string densities at given $\gamma$. The introduction of non-equilibrium sources does not change the scaling, and so we shall describe what is known from the study of thermodynamic equilibrium.

In studying the relevant scales controlling the behavior of $\bm \sigma$, one starts with equation (\ref{MainEquiEquation}). It is seen that in the region $|\lambda+\frac{1}{\pi}\log\left(\frac{\pi T}{\epsilon_F} \right)|  \sim O(1),  $which we shall refer to as the 'scaling region',  the functions $\varepsilon_m(\lambda)$ are of order $T$. This scaling region is  the region in $\lambda $ space where $\varepsilon_m(\lambda)$ varies, while outside this region $\varepsilon_m$ already approaches its asymptotic values at $\pm\infty$ depending on whether $\lambda$ is to the right or to the left of the region, respectively.
This conclusion is reached because outside the  free term (the first term on the right hand side) in equation (\ref{MainEquiEquation}) reaches its asymptotic value. Numerical studies of the solution of Eq. (\ref{MainEquiEquation}) confirm this result. These were  performed by Melnikov and reviewed in \cite{Wiegmann:Tsvelick:Review:Kondo}, but see also \cite{Desgranges:Schotte:Kondo:Numerics,Takahashi:Kondo:Numerics}. In particular, the function $\varepsilon_1(\lambda)$  diverges to $+\infty$ in the region $|\lambda|\sim O(1)$, which is to the left of the scaling region. This means that there are no holes in the $1$-string distribution in the region $|\lambda|\sim O(1)$.  A solution to the Bethe ansatz equation for $\bm \sigma$ and $\bm \sigma_h$ may then be found in which $\bm \sigma$ and $\bm \sigma_h$ approach their ground state values in the region $|\lambda|\sim O(1)$. Since in such a solution both $\sigma^{(m)}$ and $\sigma_h^{(m)}$ approach $0$ for $m>1$, the value of $\varepsilon^{(m)}(\lambda),$  for $m>1$, is irrelevant in this region, as it describes the ratio between two vanishing quantities.

\section{ Non-equilibrium Spin Sources} 
To compute the non-equilibrium source, we must have a workable representation of the overlap between the state $\<i|,$ describing the system before the quench, and any eigenstate of the interacting problem after the quench, $|E_k\>$. To do this we shall use the Slavnov overlap method, which gives the overlap between two Bethe states. Indeed the spin part of  both $\<i|$ and $|E_k\>$ are Bethe states, if we assume that $\<i|$ is an eigenstate of the free Fermion problem. The overlap between the charge part of both wavefunctions is easy to compute, since these parts of the wavefunctions are given by spinless Fermion eigenstates, and bosonization methods give the result for the overlap. 

Since Bethe states usually come non-normalized, we must actually compute $\frac{\<i|E_k\>}{\sqrt{\<i|i\>\<E_k|E_k\>}}.$ Each term in the numerator and denominator has to be separated into spin and charge parts and computed. In fact  each of these terms gives rise to its own non-equilibrium source. Nevertheless, the charge overlaps and the  norms are, in fact, largely known. And so, we shall concentrate only on the non-equilibrium source introduced by the spin part of $\<i|E_k\>$ and comment in the end shortly on the computations of all the other non-equilibrium sources.  These include source resulting from  the charge part of the overlap, $\<i|E_k\>, $ and the sources from the charge and spin parts of the norm $\<E_k|E_k\>$. The norm $\<i|i\>$ leads to no non-equilibrium sources since it has no variation with the densities $\bm \sigma $, which only control the state $|E_k\>$. It is also a property of the spinless fermion charge wave-function, that the norm is independent of the phase shift, such that the non-equilibrium source of the charge sector associated with the norm $\<E_k|E_k\>$ may also be discarded.

To set notations we shall then write:

\begin{align}
\frac{\delta F}{\delta \bm \sigma} =\gamma( \mathcal{J}_s^{(1)}+\mathcal{J}_s^{(2)} + \mathcal{J}_c),\label{varFismathcalJ}
\end{align} 
where $\mathcal{J}_s^{(1)}+\mathcal{J}_s^{(2)}= \frac{\delta}{\delta \bm \sigma} \frac{|\<\bm\sigma_i|\bm \sigma\>|^2}{\<\bm\sigma|\bm{\sigma}\>}$, where $|\bm\sigma_i\>$ is the spin part of the state $|i\>$ and $|\bm \sigma\>$ is the spin part of the state $|E_k\>$. The charge source $\mathcal{J}_c$ will be given below. Similarly the way  the spin source is distributed between $\mathcal{J}_s^{(1)}$ and $\mathcal{J}_s^{(2)}$, which is purely a question of convenience, will be described later. 

\subsection{Slavnov Overlaps}  

We shall say that the set $\bm a$ satisfies the Bethe ansatz equations with the vector $\bm b$ if for any $a_i \in \bm a$
\begin{align}
\frac{Q_{\bm b}(a_i)Q_{\bm a}(a_i+\imath)}{Q_{\bm b}(a_i + \imath)Q_{\bm a}(a_i-\imath)} =-1, \label{Betaab}  
\end{align}
where here and below, for any set $\bm d,$ we denote the monomial polynomial with zeros at the set $\bm d$ as $Q_{\bm d}$: 
\begin{align}
Q_{\bm d}(x) = \prod_{d_i\in \bm d} (x-d_i).
\end{align} 

Note that Equation (\ref{Betaab}) is just  a re-writing of (\ref{BetaLambda}) in a manner which will be convenient for our current purposes. 

The spin part of $|E_k\>$ is described by a set of rapidities  $\bm u $ that satisfy the Bethe equations with the vector $\bm z$ given by 
\begin{align}
\bm z =\Bigl(\frac{\imath}{2} -\frac{1}{g},\overset{\times N}{\overbrace{\frac{\imath}{2},\dots,\frac{\imath}{2}}}\Bigr). 
\end{align}
We shall denote this spin-state by $|\bm u\>_{\bm z},$ to explicitly denote the inhomogeneities, or simply  $|\bm u\>$. 

The spin part of the state $|i\>$ is described by a set of rapidities, $\bm v',$ satisfying the  free Fermion Bethe ansatz. Namely the set of inhomogeneities is given by:
\begin{align}
\bm z'=\Bigl(\overset{\times N}{\overbrace{\frac{\imath}{2},\dots,\frac{\imath}{2}}}\Bigr).
\end{align}
In addition a single impurity, decoupled from the fermions, must be included in the wave function. By symmetry to inversion of all spin components, we may assume, without loss of generality, that the impurity spin is pointing down in the $z$ direction. Thus the spin part of the $|i\>$ state is given, without loss of generality,  by $|\bm v'\>_{\bm z'}\otimes |\downarrow\>$.  The true overlap between the spin parts of $|i\>$ and  $|E_k\>$ is twice the overlap between $|\bm v'\>_{\bm z'}\otimes |\downarrow\>$ and $|\bm u\>_{\bm z}$, because the spin of the impurity in the state $|i\>$ can either point or down, but the overlap is equal for both possibilities, by  spin inversion symmetry. 

It may be shown by inspecting (\ref{BigPsi}) and (\ref{BigF}) that:\begin{align}
|\bm v'\>_{\bm z'}\otimes|\downarrow\> =\label{Limiting1}\lim_{\delta \to 0}\delta  \sum_l \left|\bm v'\cup\{-g_I^{-1}+\frac{\imath}{2}+\delta\}\right\>_{\bm z},
\end{align}
where $|\bm a\>_{\bm b}$ is the Bethe state with rapidities $\bm a$ and inhomogeneities $\bm b$. We thus have:
\begin{align}
\big(_{\bm z'}\<\bm v'|\otimes\<\downarrow|\big)|\bm u\>_{\bm z}=\label{Limiting2}\lim_{\delta \to 0 }\delta \,_{\bm z}\left\<\bm v^*\left|\right. \bm u\right\>_{\bm z},
\end{align}
where
\begin{align}
\bm v ^*= \bm v'\cup\{-g_I^{-1}+\frac{\imath}{2}+\delta\}. 
\end{align}
Note that here the star denotes complex conjugation, and it is just a matter of later convenience that we prefer to define the set $\bm v$ as the complex conjugate of the set of rapidities which define the initial state. 

The Slavnov formula gives and expression for the  overlap of Bethe states with two sets of $N$ rapidities $\bm u$ and $\bm v^*$, where it is assumed that $\bm u$ satisfies the Bethe equations with inhomogeneities $\bm z$. Kostov and Matsuo\cite{Kostov:Inner:Product:Domain:Wall} gave a symmetric representation for this expression, we give it here in the version presented in \cite{Kostov:Bettelheim:SemiClassical:XXX}:
\begin{align}
\<\bm v^*|\bm u\>=\prod_i\left(-Q_{\bm z}(v_i+\imath)Q_{\bm z}(u_i)\right)\lim_{\delta\to0} \delta\det B,\label{OverlapAsDet} 
\end{align}     
where $\bm c=\bm u \cup \bm v=\bm c\cup\bm v'\cup\{-g_I^{-1}-\frac{\imath}{2}+\delta\}$ and $B$ is a $2N$ by $2N$ matrix: 
\begin{align}
B_{i,j} =\delta_{i,j} -\frac{Q_{\bm z}(c_i)Q_{\bm c}(c_i+\imath)}{Q_{\bm c}(c_i)Q_{\bm z}(c_i+\imath)} \frac{1}{c_i-c_j+\imath}.
\end{align}

\section{Integral Equations for the Overlap}

The main task is to compute the source term that arises from the overlap (\ref{OverlapAsDet}), namely the expression $2{\rm Re}\frac{\delta}{\delta \bm \sigma} \log \<\bm v^*|\bm u\>$. This requires finding the variation, with respect to $\bm \sigma$, of the logarithm of the determinant of $B$ and taking the real value.  It turns out that the source term associated with the prefactor in Eq. (\ref{OverlapAsDet}), mamely the term $\prod_i\left(-Q_{\bm z}(v_i+\imath)Q_{\bm z}(u_i)\right)$, cancels out upon subtraction  by the source term associated with  $|\<\bm u|\bm u\>|$, and thus may be ignored.
We thus define $\mathcal{J}^{(1)}_s$, which was introduced in (\ref{varFismathcalJ}), as:
\begin{align}
\mathcal{J}_s^{(1)} =2{\rm Re}  \frac{\delta}{\delta \bm \sigma } \log \det B,\label{mathCalJs1}
\end{align} 
while $\mathcal{J}_s^{(2)}$ will be a similar expression involving the norm of the Slavnov matrix associated with the norm $\<\bm u|\bm u\>$.

In this section, we develop a method which will allow to compute the source term associated with the overlap by solving linear integral equations. To do so, we first represent the matrix $B$ as an operator in a certain function space. This is done in the following subsection.

\subsection{Functional Representation of the Matrix $B$}We shall use a function space representation of  the matrix $B$. In the function space representation,  any vector,  $\vec \psi$, the elements of which are given by $\psi_i$, is represented by a function $\psi(x)$ given by $\psi(x)= \sum_{v\in\bm c} \frac{\psi_i}{x-c_i}$. Thus, $\psi_i = \Res{x\to c_i} \psi(x).$ Namely, the vector $\psi_i$ is encoded in the residues of $\psi(x)$. The function,  $\psi(x),$ thus has jump discontinuities across the lines $\mathds{R} + \frac{\imath}{2}n$, with $n\in \mathds{Z}$.  

We wish to write the functional representation of $B \vec \psi$ given the function representation of $\vec \psi$. We write this as $\mathcal{B}\psi$, namely, $\mathcal{B}\psi =\sum_i \frac{(B\psi)_i}{x-c_i}.  $  The operator $\mathcal{B}$  was introduced in \cite{Kostov:Bettelheim:SemiClassical:XXX} and found to be given by
\begin{align}
\mathcal{B} = 1-\tilde{ \mathcal{P}} \frac{Q_{\bm z}(x)}{Q_{\bm u}(x)} e^{\imath \partial} \frac{Q_{\bm u}(x)}{Q_{\bm z}(x)},\label{MathcalBDef}
\end{align}  
where $\tilde{\mathcal{P}}$ is a projection operator projecting onto functions having only singularities on the set $\bm c$:
\begin{align}
(\tilde{\mathcal{P}} f )(x) =\oint_{\bm u} \frac{f(x')}{x-x'} \frac{dx'}{2\pi\imath}.
\end{align}
Here the integral should be taken around the set $\bm u$, leaving out the points $\bm z -\imath$. To prove that Eq. (\ref{MathcalBDef}) gives the correct functional representation is an easy task. Indeed, evaluating the residue of $\mathcal{B}\psi$ at $c_i$ gives $(B\psi)_i$ as required. In addition, the function $\mathcal{B}\psi$ has no other singularities other than the poles at $c_i$ showing the equality $\mathcal{B}\psi =\sum_i \frac{(B\psi)_i}{x-c_i}$. 

The operator $\mathcal{B}$ satisfies:
\begin{align}
\Res{x\to c_i}\mathcal{B}\frac{1}{x-c_j} = B_{i,j}.
\end{align}
This formula is correct before the limit $\delta\to0$ is taken. This limit corresponds  to taking    $v_N=c_{2N}\to -\frac{\imath}{2} - \frac{1}{g_I}$, where we have designated $c_{2N}=v_N$. Taking the limit we obtain:
\begin{align}
\Res{x\to c_{2N}} (x-c_{2N})\mathcal{B}\frac{1}{x-c_j}=\lim_{\delta \to 0}\delta\cdot B_{2N,j}.
\end{align} 

\subsection{Integral Equations for the inverse of $\mathcal B$.}
In order to compute $\mathcal{J}^{(1)}_s$, defined (\ref{mathCalJs1}), we would need the variation of the logarithm of the determinant of $B$. This is, of course given by:
\begin{align}
\frac{\delta \log \det B}{\delta \bm \sigma } = \tr B^{-1}\frac{\delta B}{\delta \bm \sigma}.
\end{align} 
The main difficulty in computing the right hand side of this equation is inverting the matrix $B.$  We consider then the inverse operator, $\mathcal{R}$ defined by:
\begin{align}
\mathcal{B}\mathcal{R}(x,y) = \frac{1}{x-y},\label{EquationForR}
\end{align}
then the matrix elements of $B^{-1}_{i,j}$ are equal to:
\begin{align}
B^{-1}_{i,j} = \Res{x\to c_i}\mathcal{R}(x,c_j).
\end{align}
The equation for $\mathcal{R}$ is written explicitly as:
\begin{align}
\mathcal{R}(x,y) - \oint_{\bm u}\frac{Q_{\bm u}(x'+\imath)Q_{\bm z}(x')}{Q_{\bm u}(x')Q_{\bm z}(x'+\imath)} {}  \mathcal{R}(x'+\imath,y) \frac{dx'}{(x'-x)2\pi\imath} = \frac{1}{x-y}
\end{align}
The contour of integral must surround the set $\bm u$ but avoid the set $\bm z-\imath$, where the integrand has extra poles coming from $ \frac{1}{Q_{\bm z}(x'+\imath)}$. The contour must thus be drawn to surround the rapidities which lie on the lines $\mathds{R}+\frac{\imath}{2} \mathds{Z}$ but avoiding $-\frac{\imath}{2}$ and $-\frac{1}{g_I}-\frac{\imath}{2}$. It is easy to draw a contour that surrounds the rapidities but avoids the point $-\frac{\imath}{2}$, since this point is not embedded within the rapidity distribution (as was the conclusion of subsection \ref{ScalesSubSection}), but the point $-\frac{1}{g_I}-\frac{\imath}{2}$ is more tricky. We thus define the contour $\mathcal{C}$ to surrounds the rapidities, including $z_1-\imath=-\frac{1}{g_I}-\frac{\imath}{2}$, but avoid the point $-\frac{\imath}{2}$, and separate out the contribution of the point $z_1-\imath$  as follows:
\begin{align}
\mathcal{R}(x,y) +\frac{\Res{x\to z_1}(e^{\Phi(x)})}{e^{\Phi(z_1-\imath)}}\frac{\mathcal{R}(z_1,y)}{x-z_1+\imath}- \oint_{\mathcal{C}} \frac{e^{\Phi(x'+\imath)}}{e^{\Phi(x')  }} \mathcal{R}(x'+\imath,y) \frac{dx'}{(x-x')2\pi\imath} = \frac{1}{x-y},\label{Equation4RSeparated}
\end{align}
where here and below we denote:
\begin{align}
e^{\Phi(x)} = \frac{Q_{\bm u}(x)}{Q_{\bm z}(x)}.
\end{align}
As it stands, Eq. (\ref{Equation4RSeparated}) is  a closed equation for $\mathcal{R}$, but to solve it one must know $\mathcal{R}(x,y)$ at the point $x=z_1$. As will be discussed below, we can only hope to solve the equation away from the lines $\mathds{R}+\frac{\imath}{2}\mathds{Z}$. Since the point $z_1$ lies on such a line, we must write a separate equation to determine it. This equation is obtained by setting $x$ to $z_1$ in (\ref{Equation4RSeparated}):
\begin{align}
& \frac{1}{z_1-y} \label{z1EqBeforeLimit}= \mathcal{R}(z_1,y) -\imath\frac{\Res{x\to z_1}(e^{\Phi(x)})}{e^{\Phi(z_1-\imath)}}\mathcal{R}(z_1,y)- \oint_{\mathcal{C}} \frac{e^{\Phi(x'+\imath)}}{e^{\Phi(x')  }}  \frac{\mathcal{R}(x'+\imath,y)dx'}{(z_1-x')2\pi\imath} . 
\end{align} Note now that $e^{\Phi(z_1-\imath)} =\frac{Q_{\bm u+\imath}(z_1)}{Q_{\bm z+\imath}(z_1)} $ is of order $\delta$, since $c_{2N} +\imath=z_1+\delta$ . This means that the second term on the right hand side of (\ref{z1EqBeforeLimit}) diverges as $\delta\to0$ unless $\mathcal{R}(z_1,y)$ vanishes  in the limit. No other term diverges if $y\neq c_{2N}$. If the equation is to be satisfied, we must have $\lim_{\delta\to 0}\mathcal{R}(z_1,y)=0$. 
Accordingly, we define:
\begin{align}
\beta(y) =\lim_{\delta\to0}\imath\frac{\Res{x\to z_1}(e^{\Phi(x)})}{e^{\Phi(z_1-\imath)}}\mathcal{R}(z_1,y),\label{BetaDef}
\end{align}
and obtain:
\begin{align}
& \beta(y) =\frac{1}{y-z_1} - \oint_{\mathcal{C}} \frac{e^{\Phi(x'+\imath)}}{e^{\Phi(x')  }}  \frac{\mathcal{R}(x'+\imath,y)dx'}{(z_1-x')2\pi\imath} .\label{Equation4beta} 
\end{align}
In the limit $\delta\to0,$ Eq. (\ref{Equation4RSeparated}) for  $\mathcal{R}$ turns into:
\begin{align}
\mathcal{R}(x,y) -\frac{\imath \beta(y)}{x-c_{2N}}- \oint_{\mathcal{C}} \frac{e^{\Phi(x'+\imath)}}{e^{\Phi(x')  }}  \frac{\mathcal{R}(x'+\imath,y)dx'}{(x-x')2\pi\imath} = \frac{1}{x-y}.\label{Equation4R}
\end{align}  

Examining equation (\ref{Equation4R}) for $y=c_{2N}$, one obtains that a solution exists where $\mathcal{R}(x,c_{2N})=0$ and $\beta(c_{2N})=-\imath$.  Since $\lim_{\delta\to0} \mathcal{R}(x,c_{2N})=0$, it seems reasonable that we shall eventually need the next to leading order in $\delta $ of the quantity $\mathcal{R}(x,c_{2N}),$ rather than sufficing with $\mathcal{R}(x,c_{2N})=0$. Appropriately, we define:
\begin{align}
\alpha(x) = \lim_{\delta\to0} \delta^{-1}\sum_i\frac{B^{-1}_{i,2N}}{x-c_i}.
\end{align}
Given that $\mathcal{R}(x,c_{2N})$ scales as $\delta$, Eqs. (\ref{BetaDef}) and (\ref{Equation4beta}) give now:
\begin{align}
{\alpha}(z_1) \lim_{\delta\to0}\imath\delta \frac{\Res{x\to z_1}(e^{\Phi(x)})}{e^{\Phi(z_1-\imath)}} =1,\label{AlphaAtz1}
\end{align}
while (\ref{Equation4R}) becomes:
\begin{align}
\alpha(x)- \oint_{\mathcal{C}} \frac{e^{\Phi(x'+\imath)}}{e^{\Phi(x')  }}  \frac{\alpha(x'+\imath)dx'}{(x-x')2\pi\imath} =\frac{\imath}{c_{2N}-x} \oint_{\mathcal{C}} \frac{e^{\Phi(x'+\imath)}}{e^{\Phi(x')  }}  \frac{\alpha(x'+\imath)dx'}{(z_1-x')2\pi\imath}.\label{Equation4Alpha}
\end{align}

Equations (\ref{Equation4beta}), (\ref{Equation4R}), (\ref{Equation4Alpha}) and (\ref{AlphaAtz1}) constitute integral equations that determine the functional representation of $B^{-1}$. 

The integral equations at hand have the advantage that the may be formulated as equations for $\mathcal{R}$ at points that are an $O(1)$ distance away from the lines $\mathds{R}+\frac{\imath}{2}n$. In those regions the operator $\mathcal{B}$ becomes a function of the coarse grained densities $\bm \sigma$, only. Indeed $\mathcal{B}$ depends on $\bm \sigma$ only through $\Phi$, which, in turn, is the electrostatic potential produced by the rapidities, which are assigned a charge $1$, and the inhomogeneities, assigned a charge $-1$. As one moves away from the charge distribution, which lies on the lines $\mathds{R}+\frac{\imath}{2}n$, the potential becomes a function only of the low laying Fourier modes of $\bm \sigma$, the modes with wave number of order $N$, which contain the information about the exact position of the rapidities, rather than the coarse grained density of the rapidities, decay exponentially in $N$ at an $O(1)$ distance away from the charge distribution. The conclusion is that knowledge  coarse grained density is sufficient  in finding $\mathcal{R}$. This is an advantageous situation, as one cannot hope to find the exact distribution of rapidities within the thermodynamic Bethe ansatz approach, on which the current, non-equilibrium, version of the method is built. 

We should note that Eq. (\ref{AlphaAtz1}) may be written as:
\begin{align}
\oint_{z_1}\alpha(x) e^{\Phi(x)}\frac{dx}{2\pi\imath}=\imath\oint_{z_1}\frac{ e^{\Phi(x-\imath)}}{(x-z_1)^2} \frac{dx}{2\pi\imath},
\end{align}
where the contour of integration, which is denoted by $z_1$, is to surround $z_1$ but avoid $\frac{\imath}{2}$ and $z_1+\imath$. It is possible to draw such a contour on which the coarse grained approximation for $\Phi$ holds to exponential accuracy, and as such Eq. (\ref{AlphaAtz1}) is also meaningful within the coarse graining  approach. 

\subsection{Functional Form of the Variation of the Determinant of $B$}
We return to the  variation of the determinant of $B$:
\begin{align}
\frac{\delta \log \det B}{\delta \bm \sigma } = \tr B^{-1}\frac{\delta B}{\delta \bm \sigma}.
\end{align} 
Our goal now is to represent this object in terms of the solutions to the integral equation  (\ref{Equation4beta}), (\ref{Equation4R}), (\ref{Equation4Alpha}) and (\ref{AlphaAtz1}) . We shall arrive at a contour integral representation of this object, in which $B^{-1}$ is replaced by $\mathcal{R}$, the variation of $B$ is replaced by the variation of $\mathcal B$ and the trace is achieved by a contour integration. To be able to obtain this representation, we shall need the analytic properties of $\mathcal{R}(x,y)$ in terms of $y$.  

To that aim, we note the following
identity:
\begin{align}
&\mathcal{B}\mathcal{R}(x,y) =e^{-\Phi(x)}\left(1-e^{\imath\partial_x}\right) e^{\Phi(x)} \mathcal{R}(x,y)
+\sum_{l=1}^L \frac{ Q_{\bm u}(z_l)}{Q'_{\bm z}(z_l)}
\frac{e^{-\Phi(z_l-\imath )}\mathcal{R}(z_l,y)}{x-z_l+\imath },
\label{oneplusalphaK}
\end{align}
which is obtained making use of the definition of $\mathcal{B}$, Eq. (\ref{MathcalBDef}),   and taking the projection
by removing the poles explicitly, the latter being located at the
points $z_l-\imath $.  From (\ref{EquationForR}), we may replace the left hand side of (\ref{oneplusalphaK}) by $\frac{1}{x-y}, $ then multiplying both sides of the equation by $e^{-\Phi(x)} \left(1-e^{\imath\partial_x}\right)^{-1}e^{\Phi(x)},$ we obtain:
 \begin{align}
 \mathcal{R}(x,y) = e^{-\Phi(x)} \left(1-e^{\imath\partial_x}\right)^{-1}e^{\Phi(x)}
 \left(\frac{1}{x-y}- \alpha\sum_{l=1}^L e^{-\Phi(z_i-\imath )}\frac{
 Q_{\bm u}(z_l)}{Q'_{\bm z}(z_l)}\frac{\mathcal{R}(z_l,y)}{x-z_l+\imath }\right).
 \label{SelfConstCenter}
\end{align}
Then the equation may be solved self-consistently by treating
$\mathcal{R}(z_l,y)$ on the right hand side as external parameters, solving for
$\mathcal{R}(x,y)$ and then requiring that by evaluating $\mathcal{R}(x,y)$ at $x=z_l$ we
recover these same parameters.  Indeed, setting $x$ to $z_l$ in Eq.
(\ref{SelfConstCenter}) and representing $\left(1-e^{\imath\partial_x}\right)^{-1}$ as $\sum_n e^{n\imath \partial_x} $, one realizes that
only the $n=0$ term in this sum contributes, which makes the
application of the self-consistency straightforward, leading to:
\begin{align}\left(\begin{array}{c}
\mathcal{R}(z_1,y) \\
\mathcal{R}(z_2,y) \\
. \\
. \\
\mathcal{R}(z_{2N},y)
\end{array} \right)
=\tilde{B}^{-1}\left( \begin{array}{c}
\frac{1}{z_1-y} \\ \frac{1}{z_2-y} \\ . \\ . \\ \frac{1}{z_L-y}\end{array}\right),
\label{SelfConsist}
 \end{align}
with the $2N\times 2N$ matrix $\tilde B$ given by
 \begin{align}
 \tilde B_{ln} =\delta_{i,j}+ \frac{Q_{\bm z}(z_{n}-\imath
 )Q_{\bm u}(z_n)}{Q'_{\bm z}(z_n)Q_{\bm u}(z_n-\imath )}\ \frac{1}{z_l-z_n+\imath
 }.
 \end{align}
Examining (\ref{SelfConstCenter}) and (\ref{SelfConsist}), we see that, as a function of $y$,  $\mathcal{R}(x,y)$  has poles only at $y \to x+\imath n$, $n\geq0$ and at $y \to z_i$. 

We return now to the variation of the logarithm of the determinant and write:
\begin{align}
&\frac{\delta \log \det B}{\delta \bm \sigma } = \sum_{i,j} \Res{y\to c_i} \mathcal{R}(y,c_j) \frac{\delta}{\delta \bm \sigma} \Res{x\to c_j} \mathcal{B} \frac{1}{x-c_i } =\nonumber \\&= \sum_{i,j} \Res{y\to c_i} \mathcal{R}(y,c_j)  \Res{x\to c_j} \frac{1}{x+\imath-c_i }\left( \frac{\delta}{\delta \bm \sigma} +f_{\delta \bm \sigma}(c_i) \partial_x\right)\frac{Q_{\bm c}(x+\imath)Q_{\bm z}(x)}{Q_{\bm c}(x)Q_{\bm z}(x+\imath)}
\end{align}
Here we assume that a function $f_{\delta \bm \sigma}(x)$ may be defined with the properties that  $f(c_i) = \frac{\delta c_i}{\delta \bm \sigma}$ and that $f_{\delta \bm \sigma}(x)$ is analytic around each of the lines $\mathds{R}+\frac{\imath}{2} n$, with $n\in \mathds{Z}$. This is possible if, for example, the variation in the density $\bm \sigma$ is effected by displacing the rapidities $c_i$ on each of the lines $\mathds{R}+\frac{\imath}{2}n$ harmonically. Namely the case in which there exists an  infinite set of wave numbers, $k_i, $ and amplitudes $A^{(k_i)}_i$ with $i\in \mathds{N}^0$ such that:
\begin{align}
f_{\delta \bm \sigma} (x) =A^{(k_{ |i(x)| })}_{|i(x)|} \exp\left[ \imath k_{ |i(x)| }x  \right],\label{fExample}
\end{align}
where
\begin{align}
i(x) =\lceil {\rm Im} (2x)-\frac{1}{2} \rceil 
\end{align}
and $\lceil x \rceil$ is the ceiling function on $x$ (the smallest integer larger than, or equal to $x$).  Since we may span any deformation of the rapidities using this basis of plane waves, as described by Eq. (\ref{fExample}), there is no loss of generality in the assumption that $f_{\delta \bm \sigma}$ is analytic around the lines  $\mathds{R}+\frac{\imath}{2} n$. 

Taking into account the fact that $\mathcal{R}(y,x)$, as a function of $x$, has poles only at $y+n\imath$ and $\bm z$, we may write:
\begin{align}
&\frac{\delta \log \det B}{\delta \bm \sigma } \nonumber=\sum_i\Res{y\to c_i}\left[\oint_{\bm c}  \mathcal{R}(y,x)   \frac{1}{x+\imath-c_i }\left( \frac{\delta}{\delta \bm \sigma} +f_{\delta \bm \sigma}(c_i) \partial_x\right)\frac{Q_{\bm c}(x+\imath)Q_{\bm z}(x)}{Q_{\bm c}(x)Q_{\bm z}(x+\imath)} \frac{dx}{2\pi\imath} \right.- \\ &+\left. \oint_{\bm c}\mathcal{R}'(y,x)\frac{Q_{\bm c}(x+\imath)Q_{\bm z}(x)}{Q_{\bm c}(x)Q_{\bm z}(x+\imath)} \left( f_{\delta \bm \sigma}(x) -f_{\delta \bm \sigma}(y) \right)\frac{dx}{2\pi\imath}\right],
\end{align} 
where $\mathcal{R}'(y,x) = \partial_x \mathcal{R}(y,x)$.

We now define:
\begin{align}
{\bm \nabla}^{(y)}_{\delta\bm \sigma} =\frac{\delta}{\delta \bm \sigma} +f_{\delta \bm \sigma}(y) \partial_x, \quad \mathcal{P}^{(z_1)} g(x)= g(x) -\underset{{\{z_1\}\cup\{z_1-\imath\}}}\oint \frac{g(x')}{x-x'} \frac{dx'}{2\pi \imath},
\end{align}
allowing us to write the sum over the residues at the points $c_i$,  as a contour integral over $y$, as follows:
\begin{align}
&\frac{\delta \log \det B}{\delta \bm \sigma } =\label{VariationLogB}
 \oiint_{\mathcal{C}'^2}\frac{dydx}{(2\pi\imath)^2} \mathcal{P}^{(z_1)} \left[   \frac{\mathcal{R}(y,x)}{x+\imath-y } {\bm \nabla}^{(y)}_{\delta\bm \sigma}\frac{Q_{\bm c}(x+\imath)Q_{\bm z}(x)}{Q_{\bm c}(x)Q_{\bm z}(x+\imath)}
 \right.+\\&+\left( f_{\delta \bm \sigma}(x) -f_{\delta \bm \sigma}(y) \right) \left.\frac{\mathcal{R}'(y,x)}{x+\imath-y}\frac{Q_{\bm c}(x+\imath)Q_{\bm z}(x)}{Q_{\bm c}(x)Q_{\bm z}(x+\imath)} 
\right] -\nonumber\\
&-\int_0^{\infty} \imath dp \oiint_{\mathcal{\tilde C}}\frac{dydx}{(2\pi\imath)^2} \mathcal{P}^{(z_1)}   \left\{ e^{\imath p(x+\imath-y+\imath0^+)}\left[ \mathcal{R}(y,x) {\bm \nabla}^{(y)}_{\delta\bm \sigma}\frac{Q_{\bm c}(x+\imath)Q_{\bm z}(x)}{Q_{\bm c}(x)Q_{\bm z}(x+\imath)}
 \right.\right.+\nonumber\\
 &\left. \left.+\left( f_{\delta \bm \sigma}(x) -f_{\delta \bm \sigma}(y) \right)  \mathcal{R'}(y,x)\frac{Q_{\bm c}(x+\imath)Q_{\bm z}(x)}{Q_{\bm c}(x)Q_{\bm z}(x+\imath)}
 \right]\right\}+\nonumber \\
& +\left(\frac{\delta}{\delta \bm \sigma}\frac{Q_{\bm c}(z_1)Q_{\bm z}(c_{2N})}{Q'_{\bm c}(c_{2N})Q'_{\bm z}(z_1)}\right)\left[\oint_{\underset{j\neq 1}\cup \mathcal{C}_j}\frac{dy}{2\pi\imath}\frac{\alpha(y)}{z_1-y}-\int_0^\infty \imath dp \oint_{\mathcal{C}_1}\frac{dy}{2\pi\imath}\alpha(y)e^{\imath p(z_1-y+\imath0^+)}\right], \nonumber
\end{align} 
where the operator $\mathcal{P}^{(z_1)}$  acts on the coordinate $x$ rather than $y$.  The contour labelled as $\mathcal{C}'^2$ and $\tilde{\mathcal{C}}$are   defined as follows: \begin{align}
\mathcal{C}'^2= \{ \mathcal{C}_{i}\times\mathcal{C}_j|i,j\in\mathds{Z}, i\neq j+2\},\quad\tilde{\mathcal{C}}=\{ \mathcal{C}_{i+2}\times\mathcal{C}_{i}|i\in\mathds{Z}\}.
\end{align}
where $\mathcal{C}_i$ is a contour surrounding the rapidities on the line $\mathds{R}+\imath\frac{i}{2}$, and integral over the contour $\mathcal{C}_i\times \mathcal{C}_j$ denotes $y$ integration over the contour $\mathcal{C}_i$ and an $x$ integration over the contour $\mathcal{C}_j$. For the contour $\mathcal{C}_i\times\mathcal{C}_i$ in $\mathcal{C}'^2$, the $x$ contour is to be surrounded by the $y$ contour. 

To prove Eq. (\ref{VariationLogB}) consider first the first line of that equation. The $x$ integration can be performed first by picking up the residues at the  poles at $c_i,$ stemming from $Q_{\bm c}(x)$, which is present in the denominator. The residues at these points contain two contributions, the first being  $ \mathcal{R}(y,c_i)\Res{x\to c_i} \frac{\delta}{\delta \bm \sigma} \mathcal{B} \frac{1}{x+\imath-y}$ while the second being $ \mathcal{R'}(y,c_i) ( f_{\delta \bm \sigma}(x) -f_{\delta \bm \sigma}(y))\Res{x\to c_i}  \mathcal{B} \frac{1}{x+\imath-y}$. The former contribution is desirable, while the latter must be discarded. This is taken care of in the second line of (\ref{VariationLogB}). Next, the $y$ integral is performed. Since $\mathcal{R}(y,c_i)$ has poles only at the set $\bm{c}$, we obtain the desired  $\sum_{i,j} \Res{y\to c_j}\mathcal{R}(y,c_i)\Res{x\to c_i} \frac{\delta}{\delta \bm \sigma} \mathcal{B} \frac{1}{x+\imath-c_j}$. 

As for the third line of (\ref{VariationLogB}) -- there the factor $\frac{1}{x+\imath -y}$ is  represented  as $\int_0^\infty \imath dp e^{\imath p (x+\imath-y+i0^+)}$,  in order to avoid the divergence of the factor $\frac{1}{x+\imath -y},  $ which would have appeared, had this factor been included directly.  Note that the $p$ integral is performed last in order for the integral to converge. That the integral indeed converges, if performed in this order, can be seen by considering that the $x$ and $y$ integrals result in the replacement of the $p$-dependent factor by the factor $  \int_0^\infty \imath dp e^{\imath p (c_i+\imath-c_j+i0^+)}$,  associated with picking up the poles of the integrand at $c_i$ for the $x$ integral and $c_j$ for the $y$ integral. The contour $\tilde {\mathcal{C}}$ is drawn such that if  $c_i$ is on the line $\mathds{R} + \imath \frac{k}{2}$, for some $k$, then $c_j$ is on the line $\mathds{R}+\imath\frac{k+2}{2}$. This leads to the fact that  the integral $  \int_0^\infty \imath dp e^{\imath p (c_i+\imath-c_j+i0^+)}$ converges.

Finally, it should be shown that the operator $\mathcal{P}^{(z_1)}$ can be effected by macroscopic  contour integration away from  $\mathcal{R} +\frac{\imath}{2}\mathds{Z},$ even as the operator $\mathcal{P}^{(z_1)}$ is defined as a contour integral around microscopic contours around $z_1$ and $z_1-\imath$.  We demonstrate that this is possible on the example of the first integrand in (\ref{VariationLogB}), where we show that the contour integral around $z_1$ can be written in terms of macroscopic contour integrals as follows:
\begin{align}
&\oint_{\{z_1\}}\frac{\mathcal{R}(y,x')}{(x-x')(x'+\imath-y) } {\bm \nabla}^{(y)}_{\delta\bm \sigma}\frac{Q_{\bm c}(x'+\imath)Q_{\bm z}(x')}{Q_{\bm c}(x')Q_{\bm z}(x'+\imath)} \frac{dx'}{2\pi\imath}\nonumber=\\&=\frac{f_{\delta_{\bm \sigma}}(z_1)}{(x-z_1)(z_1+\imath-y)} \frac{Q'_{\bm z}(z_1)}{Q_{\bm z}(z_1+\imath)}\frac{\oint_{\mathds{R}+\frac{\imath}{2}}\frac{ Q_{\bm c}(x+\imath)}{x-z_1}}{\oint_{\mathds{R}+\frac{\imath}{2}}\frac{ Q_{\bm c}(x)}{x-z_1}} \oint_{\mathds{R}+\frac{\imath}{2}}\mathcal{R}(y,x)\frac{dx}{2\pi\imath}.
\end{align}
All further integrals going into the application of the operator $\mathcal{P}^{(z_1)}$ may be treated in an analogous fashion, by noting that the poles of $\mathcal{R}(y,x)$ as a function of $x$ are only on the $z_1$ at $y+\imath n$. The same goes to the pre-factor of the last integral in (\ref{VariationLogB}). 

Upon taking twice the real value of the right hand side, equation (\ref{VariationLogB}) is the desired equation for the non-equilibrum source associated with the overlap, $\mathcal{J}_s^{(1)}$. It contains only $\mathcal{R}$ and $ \Phi$ at points an $O(1)$ distance away from the distribution of rapidities, and, as such, contain coarse grained objects only. A subtle point is that the $p$ integration, would also pick up Fourier components  with wave numbers of order $N$\ of the integrand. If in all the objects in (\ref{VariationLogB}) we substitute the coarse grained expressions, there is a question of whether a large $p$ contribution is missed. We conjecture that this is not the case, based on the results of \cite{Kostov:Bettelheim:SemiClassical:XXX}, where an expression for the determinant of $B$\ was given in a fairly general case. The expression given there shows that the determinant depends only on coarse grained quantities. Assuming that this is the case here as well, it is appropriate to average the right hand side of  (\ref{VariationLogB}) over all configurations with the same coarse grained densities. Such averaging, which is in essence a coarse graining procedure,  is presumed to remove high wave number from the integrand in the $p$ integral. Admittedly, though, this is a weak point of our derivation. 

\subsection{The Norm}
In the previous subsections, we have treated the spin sector of the overlap $\<i|E_k\>$. The spin sector of the norm may be computed in the same method by substituting for $\bm c$ the expression $\bm c = \bm u \cup (\bm u +\varepsilon)$ and taking $\varepsilon
\to0$. In addition, the point $-\frac{1}{g_I} - \frac{\imath}{2}$ is no longer occupied by a rapidity. The result is that the integral equation that determine $\mathcal{R}$ is (\ref{Equation4RSeparated}), with Eq. (\ref{z1EqBeforeLimit}) determining $\mathcal{R}(z_1,y).$ In addition there is no function $\alpha(x)$, that need to be solved.  The  expression for $\mathcal{J}_s^{(2)}$, introduced in (\ref{varFismathcalJ}),   is then again given in   (\ref{VariationLogB}),  with  $\mathcal{R}$ being the solution of the integral equation appropriate for the norm, and  in which $\alpha$ is set to zero. In applying the operator $\mathcal{P}^{(z_1)}$ one notes a different behavior at $z_1-\imath$ as compared to what is encountered when the overlap is computed, but the general method of removing the poles at $z_1$ and $z_1-\imath$ is the same.

\subsection{Functional Derivatives}
The reader may have noticed that, in contrast to what was the case in previous chapters, the functional derivative which was taken within this section was with respect to a variation of the density $\bm \sigma$, which corresponds to a harmonic displacement of the rapidities. Indeed, in this section we gave an expression for the sources $\mathcal{J}_s^{(i)}$ which correspond to $f_{\delta \bm \sigma}(c_i) = \frac{\delta c_i}{\delta \bm \sigma}$, where $f_{\delta \bm \sigma}$ is given in (\ref{fExample}), while the derivative of the free energy was effected by $\frac{\delta}{\delta \sigma^{(m)}(\lambda)}$. Let us write $\frac{\delta}{\delta A^{(k)}_i}, $ for the  derivative describing a harmonic displacement of the rapidities, which is appropriate since $A^{(k)}_i$ denotes the amplitude of the harmonic  displacement of the rapidities, as     can be seen in Eq. (\ref{fExample}). 

It is necessary, in order to be able to apply Eq. (\ref{varFismathcalJ}), to write down the transformation between the types of derivatives. This is easily done by noting that the variation with respect to $A_i^{(k_i)}$ is a variation with respect to  $k$-th mode of the Fourier transform of the density of rapidities on the line $\mathds{R} +\frac{\imath}{2} i.$ As a result, what follows is:
\begin{align}
\frac{\delta}{\delta \sigma^{(m)}(\lambda)} =\sum_{0\leq  i\leq \frac{m}{2}}\int dke^{\imath k\lambda}  \frac{\delta}{\delta A_{m-2i}^{( k)}}.
\end{align}
We also give the inverse transformation:
\begin{align}
\frac{\delta}{\delta  A_i^{(k)}} =\int\frac{d\lambda}{2\pi}  e^{-\imath k\lambda} \left( \frac{\delta}{\delta \sigma^{(i)}(\lambda)} - \frac{\delta}{\delta \sigma^{(i-2)}(\lambda)}\right),
\end{align}
where, for $i=0,1$,   it is   implied that $\frac{\delta}{\delta \sigma^{(-2)}(\lambda)} =0, $ $ \frac{\delta}{\delta \sigma^{(-1)}(\lambda)} = 0, $ respectively.

\section{Charge Sector}

We reconsider the charge sector, which was neglected throughout the above. We assume that the charge wavefunction of the initial state, $|i\>$ of (\ref{Echo}), is a coherent state of the bosonic excitations of the the spinless fermions gas.
More explicitly, given the  following bosonic operators:
\begin{align}
J_k =\sum_j \psi^\dagger_{j-k} \psi_{j},\quad[J_k,J_{l}]=k\delta_{k,-l},
\end{align}
where $\psi_j$ is a fermionic annihilation operator with wave number $\frac{2\pi  j}{L} $.  We may thus denote the charge part of $|i\>$ by  $| \bm t\>$, with $\bm t$ a semi-infinite vector  $\bm t=(t_1,t_2,\dots),$ the elements of which denote  eigenvalues of the operator $J_k$ on the state:
\begin{align}
J_k |\bm t\> = t_k |\bm t\>, \quad \mbox{for }k>0,
\end{align}
the normalization of $|\bm t\>$ is determined by the following:
\begin{align}
|\bm t\> = e^{\sum_k \frac{t_k J_{-k} }{k} }|0\>.
\end{align}
We shall also denote the spin part of the state $|i\>$ by $|\bm \sigma_i\>$, such that we have:
\begin{align}
|i\> = |\bm \sigma_i\>\otimes|\bm t\>e^{-\frac{\bm t^\dagger\Gamma \bm t}{2} }\label{iChargeSpin},
\end{align}
where $\Gamma$ is the diagonal matrix $\Gamma_{k,l} = k \delta_{kl}$.
The factor $e^{-\frac{\bm t^\dagger\Gamma \bm t}{2} }$ is necessary to normalize the coherent state.

The charge part of the state $|E_k\>$ in (\ref{Echo}) is described by spinless fermions which pick up  a phase  of $\delta(\bm \sigma)$, the latter being given in Eq. (\ref{phaseShift}), going around the ring on which the system is defined. 
The eigenstates of the charge sectors may thus be described by occupation numbers, $n_j$, with $n_j\in\{0,1\}$, describing the occupation of the single particle state with momentum $ \frac{\hbar}{2\pi L}\left( j +\frac{\delta(\bm \sigma)}{2\pi} \right)$  and energy $\Delta\left( j + \frac{\delta(\bm \sigma)}{2\pi}\right)$, where $\Delta =\frac{\hbar v_F }{2\pi L}. $ We shall denote such a state by $|\bm n , \delta (\bm \sigma)\>$, such that we have:
\begin{align}
|E_k\> = |\bm \sigma\> \otimes|\bm n,\delta(\bm \sigma)\>.\label{EkChargeSpin}
\end{align}
Separating in (\ref{Echo}) the charge and spin sectors,  making use of (\ref{iChargeSpin}) and (\ref{EkChargeSpin}), we obtain:
\begin{align}
&L(t) = \sum_{\bm \sigma,\bm n}|\<\bm \sigma_i | \bm \sigma\>|^2 |\<\bm t|\bm n,\delta(\bm \sigma)\>|^2e^{-\gamma( E(\bm \sigma) + \Delta \sum_j j n_j)-\bm t^\dagger \Gamma \bm t   }  
\end{align}
It is known from the theory of free fermi liquids, and the bosonization thereof, that one may write 
\begin{align}
 |\bm n , \delta (\bm \sigma)\> =e^{-\frac{ \delta(\bm\sigma)\varphi(0)}{2\pi}} |\bm n,0\>.
\end{align}
where $\varphi(\gamma) = \sum_{k\neq0}\frac{ J_k}{k} e^{\Delta k \gamma}.$ This leads to the following, bosonized, form of the Loschmidt echo:
\begin{align}
&L(t)=\sum_{\bm \sigma}|\<\bm \sigma_i | \bm \sigma\>|^2 e^{-\gamma E(\bm \sigma)   }\<\bm t|e^{-\frac{\delta(\bm \sigma)\varphi(0)}{2\pi}} e^{-\gamma H^{(0)}} e^{\frac{\delta(\bm \sigma)\varphi(0)}{2\pi}} |\bm t\>e^{-\bm t^\dagger\Gamma \bm t }  .\nonumber \end{align}
This also can be written as
\begin{align}
L(t)=\sum_{\bm \sigma}|\<\bm \sigma_i | \bm \sigma\>|^2 e^{-\gamma E(\bm \sigma)   }\<\bm t(-\gamma)|e^{-\frac{\delta(\bm \sigma)\varphi(-\gamma)}{2\pi}}  e^{\frac{\delta(\bm \sigma)\varphi(0)}{2\pi}} |\bm t\>e^{-\bm t^\dagger\Gamma \bm t },\label{LtBosonized}
\end{align}
where $\bm t(\gamma) $ is a vector, the $k$-th element of which is $t_k e^{-\gamma \Delta k}.$ The expression on the right hand side of (\ref{LtBosonized}) can be easily computed making use of the commutation relations of the bosonic operators, yielding:
\begin{align} 
L(t)\label{LtWithChargeFinal}=\sum_{\bm \sigma}|\<\bm \sigma_i | \bm \sigma\>|^2 e^{-\gamma E(\bm \sigma)   }\frac{e^{\sum_{k>0}\left(1-e^{-\gamma  \Delta k}\right) \left(\frac{\delta(\bm \sigma)}{\pi}{\rm Re}(t_k)-k|t_k|^2 \right)  }}{(1-e^{-\gamma \Delta k})^{\frac{\delta(\bm \sigma)^2}{4\pi^2}}}
\end{align}

The factor $ e^{-\gamma E(\bm \sigma)   }$ in (\ref{LtWithChargeFinal}) and the measure of integration going from the sum to an integration over $\bm \sigma$, is taken into account in the free energy. The factor $|\<\bm \sigma_i | \bm \sigma\>|^2$ has been the subject of most of this paper. We are left with the last two factor in (\ref{LtWithChargeFinal}), which were not taken into account until now and represent the charge non-equilibrium sources.  The sources are obtained by taking the logarithm of these two factors and then taking a variation with respect to $\sigma^{(m)}(\lambda)$. We obtain the following charge sources, which we denote by $\mathcal{J}_c$:
\begin{align}
\mathcal{J}_c=\Theta_m(\lambda)\sum_{k>0}\left(-\frac{\delta (\bm \sigma)}{2\pi^2} \log( 1-e^{-\gamma \Delta k} ) +\frac{1}{\pi} \sum_{k>0}(1-e^{-\gamma  \Delta k}){\rm Re}(t_k)\right) .
\end{align}

\section{Conclusion}
We presented integral equation to determine the non-equilibrium sources which appear in the Bethe ansatz equations when one considers the Loschmidt echo. We have obtained linear integral equations, through the solution of which the non-equilibrium source may be determined. It would be interesting to understand whether the equations may be solved numerically, or alternatively, whether asymptotes of the generating function could be found by analytical means, by examining the appropriate limits of the integral equations. For example, the large and small $\gamma$ limit could be examined. 

In addition to the object that we have studied here, namely the Loschmidt echo, other, more complicate objects could be treated with the same general approach. It seems that as one moves away from the relatively simple object that was dealt with here, the treatment of the problem within the current approach becomes rapidly more cumbersome. Nevertheless, and at the same time,  it seems that the overall scheme of solving the problem remains largely the same. Indeed, many non-equilibrium objects, such as the current at  intermediate voltages, may be computed by considering setups in which one starts with some free fermion initial state, turns on the interaction, waits until a steady state establishes and computes the desired observable, such as the current\cite{Andrei:Mehta:Bethe:Ansatz:Open:Systems}. The current paper demonstrates that starting with a free fermion state and allowing it to evolve, albeit in imaginary time, may be treated using the Bethe ansatz method, by treating the overlaps making use of the Slavnov determinant expression for such overlaps. Computing then observables, can then be achieved using the same Slavnov determinant approach, by representing the observable, using the inverse scattering method, with operators burrowed from the algebraic Bethe ansatz. Once this is done, observables can be computed using the Slavnov determinant method. 

It thus seems that the current approach may provide a promising avenue to study different non-equilibrium problems. It may also be said that the approach is likely to be tractable also in applications to other systems other than the isotropic Kondo problem. In this respect, a treatment of the anistropic Kondo problem, using a different, but perhaps related approach can be found in Ref. \cite{Saleur:Lukyanov:Overlaps:Kondo,Saleur:Corssover:In:Impurity}. All these issues are of course beyond the scope of the current paper. 

\section{Acknowledgement}
I wish to thank I. Kostov for  many detailed discussions. I\ would also like to thank D. Serban, P. Wiegmann, N. Andrei and Y. Vinkler-Aviv for their input. This work was supported financially by the Israel Science Foundation (Grant No. 852/11) and by the Binational Science Foundation (Grant
No. 2010345). 

\appendix

\bibliographystyle{unsrt}

\bibliography{mybib}

\end{document}